# A PDMS-based broadband acoustic impedance matched material for underwater applications

R.-M. Guillermic, M. Lanoy, A. Strybulevych, J.H. Page

Department of Physics & Astronomy, University of Manitoba, Winnipeg, Manitoba, Canada R3T 2N2

**Abstract**

Having a material that is matched in acoustic impedance with the surrounding medium is a considerable asset for many underwater acoustic applications.  In this work, impedance matching is achieved by dispersing small, deeply subwavelength sized particles in a soft matrix, and the appropriate concentration is determined with the help of Coherent Potential Approximation and Waterman & Truell models.  We show experimentally the validity of the models using mixtures of Polydimethylsiloxane (PDMS) and $TiO_2$ particles.  The optimized composite material has the same longitudinal acoustic impedance as water and therefore the acoustic reflection coefficient is essentially zero over a wide range of frequencies (0.5 to 6 MHz).  PDMS-based materials can be cured in a mold to achieve desired sample shape, which makes them very easy to handle and to use.  Various applications can be envisioned, such the use of impedance-matched PDMS in the design and fabrication of acoustically transparent cells for samples, perfectly matched layers for ultrasonic experiments, or superabsorbing metamaterials for water-borne acoustic waves.

## 1. Introduction

When an ultrasonic wave propagates from a medium of given impedance to another one with a different impedance, it is partially reflected, causing transmitted signal loss and echo generation. In cases where the best possible transmission through an interface is wanted, matching the impedance of both materials is crucial. The concept of impedance matching is the heart of transducer design, as these devices contain a matching layer adapted to optimize transmission into the medium that is being studied [1,2]. Impedance matched materials are also very useful to make sample confinement cells, since otherwise their walls tend to introduce multiple reflections and to degrade the transmission. Avoiding reflections is also an important aspect in the design of acoustic metamaterials, such as super absorbers [3] or flat lenses [4], which have shown very interesting features but are often ill-matched to the surrounding propagation medium.

Acoustic impedance is defined as

$$Z = \rho \frac{\omega}{k} = \frac{\rho v}{1 + i \frac{\alpha v}{2\omega}}, \qquad (1)$$

where $\rho$ is the density of the material and $k$ is the wave number defined as $k = \omega/v + i\,\alpha/2$, with $v$ the longitudinal phase velocity, $\omega$ the angular frequency and $\alpha$ the (intensity) attenuation coefficient. When $\alpha \ll 2\omega/v$, equation (1) simplifies to $Z = \rho v$. Propagation of sound waves through the interface between two media is schematically shown in Figure 1. The amplitude transmission coefficient from a medium 1 to a medium 2 can be written

$$t = \frac{2Z_2}{Z_1 + Z_2} \qquad (2)$$

and the reflection coefficient is

$$r = \frac{Z_2 - Z_1}{Z_2 + Z_1}, \tag{3}$$

with $Z_1$ and $Z_2$ being the impedances of medium 1 and 2 respectively. If $Z_1 = Z_2$, transmission is perfect and reflection cancels (Figure 1b).

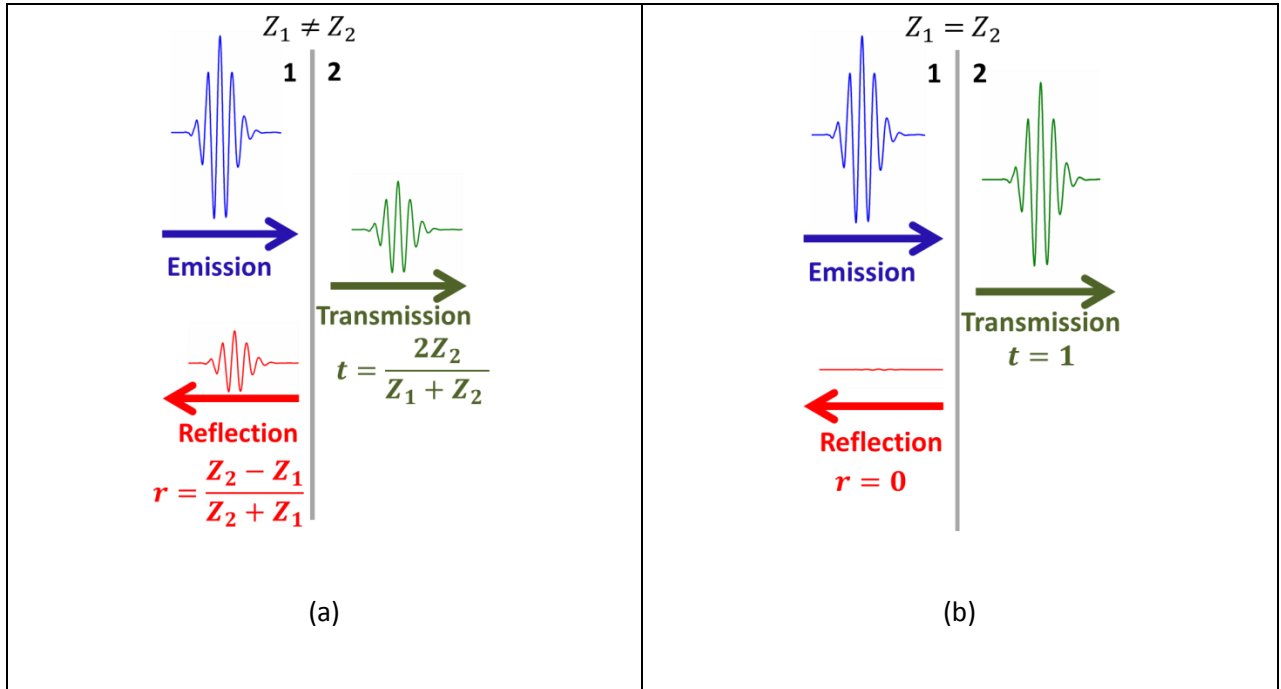

**Fig. 1.** Wave propagation through an interface between two media. If the impedance is matched, $Z_1 = Z_2$, no reflection occurs and the wave is fully transmitted.

In this article, we demonstrate the possibility of achieving impedance matching between a fluid and a viscoelastic matrix by incorporating a population of small scatterers to the matrix. We provide an experimental illustration of the technique by tackling the case of Polydimethylsiloxane (PDMS), and show that its impedance can be matched to water when the PDMS is loaded with the appropriate amount of titanium dioxide particles ($TiO_2$). In a composite medium composed of particles dispersed in a matrix, the wave is multiply scattered off the particles and the description of the sound propagation can become complex. However, in the long wavelength regime, one can adopt an effective medium approach and describe the sample as a homogeneous medium with modified wave properties. For example, the Coherent Potential Approximation (CPA) model can be used to calculate the properties of this type of composite material for frequencies far below the resonant frequencies of the filler particles and thus predict the concentration that will lead to impedance matching. Other scattering theories, such as the Waterman & Truell model, can also be used to calculate the effective wave vector of the composite material through the calculation of monopolar and dipolar scattering functions of individual scatterers.

**2. Theory and Model predictions**

In this section, we compare two models in order to predict the acoustic effective properties of composite materials and apply them to the case of submicron $TiO_2$ particles dispersed in PDMS.

*2.1. Coherent Potential Approximation*

In the case where the wavelength is larger than the individual particles, effective medium theories, such as the Coherent Potential Approximation model [5], can be used. For elastic composites with a three-dimensional dispersion microstructure, the effective shear and longitudinal complex moduli of the medium (respectively $G_{\text{eff}} = G'_{\text{eff}} + iG''_{\text{eff}}$ and $M_{\text{eff}} = M'_{\text{eff}} + iM''_{\text{eff}}$) can be calculated from the elastic moduli of its two components. We have:

$$M_{\text{eff}} = \frac{M_0[4(G_0 - G_1) + 3M_1]}{3pM_0 + (1-p)[4(G_0 - G_1) + 3M_1]} - \frac{4}{3}(G_0 - G_{\text{eff}}) \quad (4)$$

$$A\left(\frac{M_{\text{eff}}}{M_0}\right)^2 + B\left(\frac{M_{\text{eff}}}{M_0}\right) + C = 0, \quad (5)$$

where subscripts 1 and 0 represent the dispersed phase (particles) and the matrix (host medium) respectively, and $p$ is the volume fraction of fillers [5-7]. $A$, $B$ and $C$ are functions of $M_0$, $M_1$, $G_0$ and $G_1$ [5,8].

The effective density $\rho_{\text{eff}}$ can be found simply with the mixture law:

$$\rho_{\text{eff}} = \rho_0(1-p) + \rho_1 p \tag{6}$$

The effective longitudinal phase velocity $v_{\text{eff}}$ can be calculated using the relationships between the longitudinal modulus, velocity and attenuation:

$$M'_{\text{eff}} = \rho_{\text{eff}} v_{\text{eff}}^2 \frac{1 - \left(\frac{\alpha v_{\text{eff}}}{2\omega}\right)^2}{\left[1 + \left(\frac{\alpha v_{\text{eff}}}{2\omega}\right)^2\right]^2} \tag{7}$$

$$M''_{\text{eff}} = \frac{\rho_{\text{eff}} v_{\text{eff}}^2 \left(\frac{\alpha v_{\text{eff}}}{2\omega}\right)}{\left[1 + \left(\frac{\alpha v_{\text{eff}}}{2\omega}\right)^2\right]^2} \tag{8}$$

If the approximation of low attenuation is made ($\alpha \ll 2\omega/v_{\text{eff}}$), then we obtain

$$v_{\text{eff}} = \sqrt{\frac{M'_{\text{eff}}}{\rho_{\text{eff}}}}, \tag{9}$$

and the effective impedance is then

$$Z_{\text{eff}} = \rho_{\text{eff}} v_{\text{eff}} = \sqrt{M'_{\text{eff}} \rho_{\text{eff}}}. \tag{10}$$

*2.2. Waterman & Truell Model*

In the Waterman & Truell multiple scattering model [9], the propagation of the ensemble-averaged wave field in a heterogeneous medium is shown to be characterized by an effective wavevector $k_{\text{eff}}$ that is determined by the number of scattering particles per unit volume $n$ and the far field scattering amplitude $f(\theta)$ of a single scatterer:

$$\left(\frac{k_{\text{eff}}}{k_0}\right)^2 = \left[1 + \frac{2\pi n f(0)}{k_0^2}\right]^2 - \left[\frac{2\pi n f(\pi)}{k_0^2}\right]^2 \qquad (11)$$

Here $k_0$ is the matrix wave vector and $f(0)$, $f(\pi)$ are the far field forward scattered and backscattered amplitudes of a single particle. The particle positions are assumed to be uncorrelated. At long wavelengths compared with the particle size, it is sufficient to account only for the monopolar and dipolar contributions, $f_0$ and $f_1$, to the scattering amplitude $f(\theta)$. Then, it is straightforward to show that $f(0) = f_0 + f_1$ and $f(\pi) = f_0 - f_1$, allowing equation (11) to be written as the product of two factors that can be identified with the effective modulus and density:

$$\frac{M_0}{M_{\text{eff}}} = 1 + \frac{4\pi n f_0}{k_0^2} \qquad (12)$$

$$\frac{\rho_{\text{eff}}}{\rho_0} = 1 + \frac{4\pi n f_1}{k_0^2} \qquad (13)$$

Here the monopolar and dipolar scattering functions $f_0$ and $f_1$ can be calculated with the Epstein & Carhart and Allegra & Hawley model (ECAH) [10-11], and $M_0$ and $\rho_0$ are the matrix modulus and density respectively. The effective velocity and impedance can then be obtained using equations (9) and (10). In the case considered in this paper, particles are very small (< 100 nm) compared to the wavelength (between 0.15 and 2 mm), so that the scattering is weak, no resonance will occur within the range of frequencies investigated, and these effective medium models are expected to give an accurate description of wave propagation.

*2.3. Application of the models to PDMS-TiO$_2$ composites*

The properties of the TiO$_2$ particles and the PDMS matrix at 500 kHz are summarized in Table 1. The PDMS properties were measured with longitudinal transmission experiments (see materials and methods section) and shear reflection experiments (as described in references [12-13]). Properties of TiO$_2$ (rutil) were obtained from [14].

| Parameters | TiO$_2$ (1) | PDMS (0) |
|---|---|---|
| $M'$ (Pa) | 3.69x10$^{11}$ | 1.06x10$^9$ |
| $M''$ (Pa) | 1.0x10$^6$ | 4.63x10$^6$ |
| $G'$ (Pa) | 1.15x10$^{11}$ | 1.0x10$^6$ |
| $G''$ (Pa) | 1.0x10$^6$ | 0.6x10$^6$ |
| $\rho$ (kg.m$^{-3}$) | 4260 | 1020 |
| Radius $R$ (nm) | < 100 | |

**Table 1.** Properties of TiO$_2$ particles [14] and PDMS, used in the CPA and Waterman & Truell model predictions. PDMS parameters of our samples in this table were determined experimentally at 500 kHz. Numbers 1 and 0 represent the dispersed phase (particles) and the matrix respectively.

Values from Table 1 were used to calculate phase velocity and density with both the Coherent Potential Approximation and Waterman & Truell models. Figure 2 shows the predicted phase velocity and density as a function of volume fraction of TiO$_2$ particles using the parameters from Table 1. Both models agree with each other very convincingly within the entire explored density range. The density is identical for both models which is not surprising as the frequency used for the experimental characterization of PDMS (500 kHz) is far from the resonance frequency of an individual TiO$_2$ particle (which has a radius smaller than 100 nm). From these predictions, we can deduce an optimum volume fraction of particles to mix with PDMS in order to have perfect impedance matching with a range of other materials. This value depends of the properties of the surrounding medium. For water at ambient temperature having an impedance of 1.50 MRay, a volume fraction of 20.6 % of TiO$_2$ particles dispersed in a PDMS matrix is predicted to give a very good impedance match, with a velocity of 889 m.s$^{-1}$ and a density of 1690 kg.m$^{-3}$.

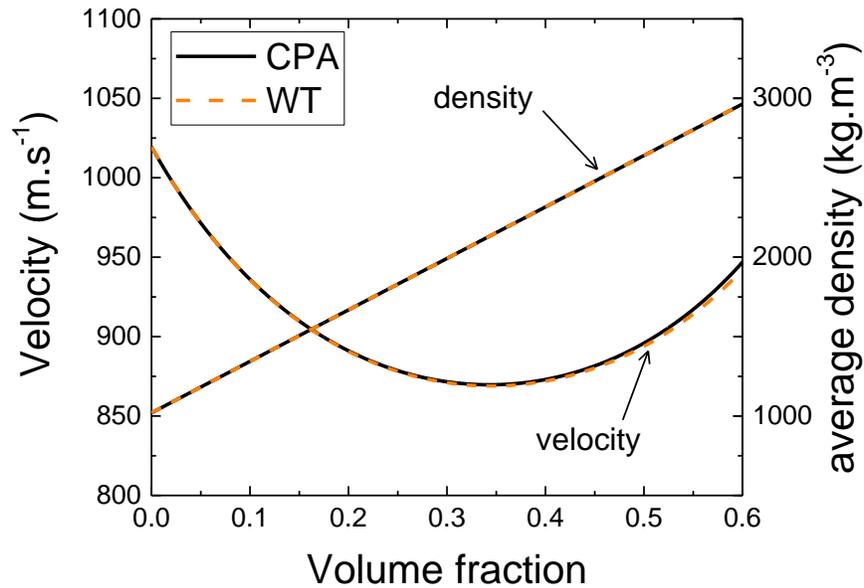

**Fig. 2.** Velocity and density of composite PDMS-TiO$_2$ materials as a function of volume fraction of TiO$_2$ particles. Predictions, calculated with PDMS parameters measured at 500 kHz, are shown for the Coherent Potential Approximation (CPA) and Waterman & Truell (WT) models.

## 3. Materials and Methods

*3.1. Sample preparation*

The samples consist of a PDMS matrix with a dispersion of TiO$_2$ particles. We used RTV 615 (GE Silicones) with a monomer/cross-linker ratio of 20:1, and TiO$_2$ particles, Titanium(IV) oxide ≥99 %, from Sigma-Aldrich. The experimental protocol was the same for all the samples studied. The monomer part of the PDMS was poured into a 250 mL beaker, then the appropriate amount of TiO$_2$ was added and mixed thoroughly with a spatula. The mix was placed in a vacuum chamber to remove all air inclusions and in an ultrasonic bath to break up the particle agglomerates. If needed, the process was repeated many times and the mix was allowed to rest for 2 to 3 days to optimize the homogeneity of the samples.

Then, the cross-linker was added, the sample was mixed and poured onto a flat circular mold. Finally, the mixture was evacuated again in the vacuum chamber to remove all remaining air inclusions and laid flat in an oven at 90 °C for 120 min. The sample was then removed from the mold and the thickness was measured (typically between 1.3 and 4.3 mm). The density of each PDMS-TiO$_2$ sample was measured using a specific gravity bottle. Samples with concentrations of TiO$_2$ particles ranging from 10 % to 25 % of the total volume were prepared.

*3.2. Longitudinal acoustic experiments*

Pairs of immersion "plane wave" ultrasonic transducers from Panametrics, of central frequencies ranging from 500 kHz to 5 MHz, were placed face-to-face and parallel to each other in a water tank (Figure 3). The different samples were positioned between both transducers using a vertical holder previously covered with Teflon$^{TM}$ tape (to prevent signal transmission through the holder) and the time-dependent transmission and reflection signals were acquired, together with a reference signal. The reference was measured with the sample removed, without changing the setup (holder still in place and transducers at the same position).

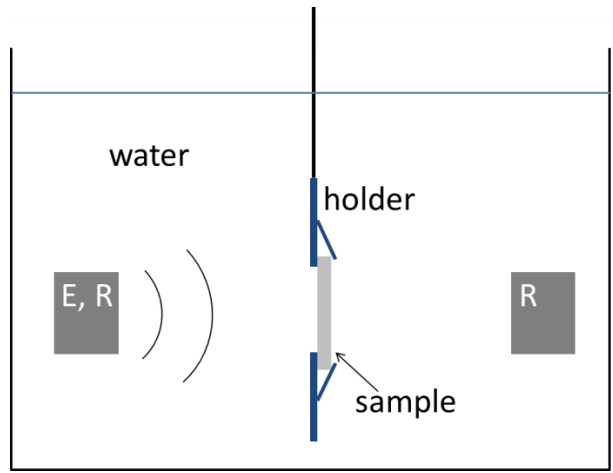

**Fig. 3.** Experimental setup for longitudinal acoustical measurements.

We obtained the frequency dependence of the transmitted amplitude and phase by performing pulsed experiments, and analyzing the data with a Fourier analysis method [15]. From the ratio of the fast Fourier transforms $\text{FFT}_\text{sample}/\text{FFT}_\text{reference}$, which is equal to $(A_\text{sample}/A_\text{reference})\exp(i\Delta\phi)$, one can easily extract the attenuation (from the amplitude ratio $A_\text{sample}/A_\text{reference}$) and phase velocity (from the phase difference $\Delta\phi$, taking care to account for the phase shift in the reference due to transport through a water medium of the same thickness as the sample). Since the model calculations showed that the dynamic mass density is equal to the static (zero frequency) average density [Equ. (6)], the impedance could be accurately calculated from the measured density and these acoustic parameters.

## 4. Results and Discussion

The ultrasonic velocity and attenuation of the PDMS-TiO$_2$ samples were measured in two frequency bands (400-kHz – 1.3 MHz, and 3 – 6 MHz), thereby spanning the range of frequencies commonly encountered in immersion ultrasonic experiments and applications. Four TiO$_2$ particle concentrations, 10 %, 17.5 %, 19.2 % and 25 %, were investigated, and compared with pure PDMS. As predicted by the models discussed in section 2, the velocities for PDMS-TiO$_2$ with these concentrations of particles are all lower than pure PDMS, and are quite close together in magnitude. Unlike the velocity, the attenuation varies quite strongly with frequency, and becomes appreciable at the higher frequencies for the larger concentrations of TiO$_2$ particles[1]. Nonetheless, the attenuation is still sufficiently low throughout the frequency range investigated that the factor $\alpha v/2\omega$ is much less that 1 (typically ~0.02), so that it is an excellent approximation [16] to use the simple relationships between velocity, modulus and impedance given by Equations (9) and (10). The experimentally measured values of the density of the samples are in excellent agreement with the theoretical predictions given by Equs. (6) and (12).

---

[1] More details on the attenuation are given at the end of this section.

The impedance was determined from the measured velocities and densities for pure PDMS and for PDMS-TiO$_2$ with the four particle concentrations. Figure 4a reports the measured values at 500 kHz along with the corresponding model predictions. Both models successfully capture the experimentally observed behaviour, although the experimental values are slightly higher than the theoretical predictions. This deviation, which is most apparent at the higher concentrations, is due to a very weak frequency dependence of the velocity due to the viscoelastic properties of PDMS – behaviour that is not accounted for in either model. As a consequence, the concentration required for achieving impedance matching to water at 20 °C ($Z = 1.48$ MRay [17]) is slightly less in practice (17.5 %) than predicted by the models (19.3 %). Also, at the experimental impedance matching volume fraction (17.5 %), the models predict an impedance of 1.43 MRay. These measurements having been obtained at 500 kHz, the question arises whether these results hold for different driving frequencies. As shown by Figure 4b, the frequency dependence of the impedance is almost negligible (with a variation that is typically of order 2%, and always less than 4% even at the highest concentration) from 400 kHz to 6 MHz. This suggests that the impedance matching is likely to be broadband. For example, in the case of the 17.5 % TiO$_2$ concentration, we measured an impedance of 1.47 MRay at 500 kHz and 1.51 MRay at 5 MHz, values that are very close to the water impedance at 20 °C. To demonstrate this behaviour directly, we now investigate the case of a broadband excitation.

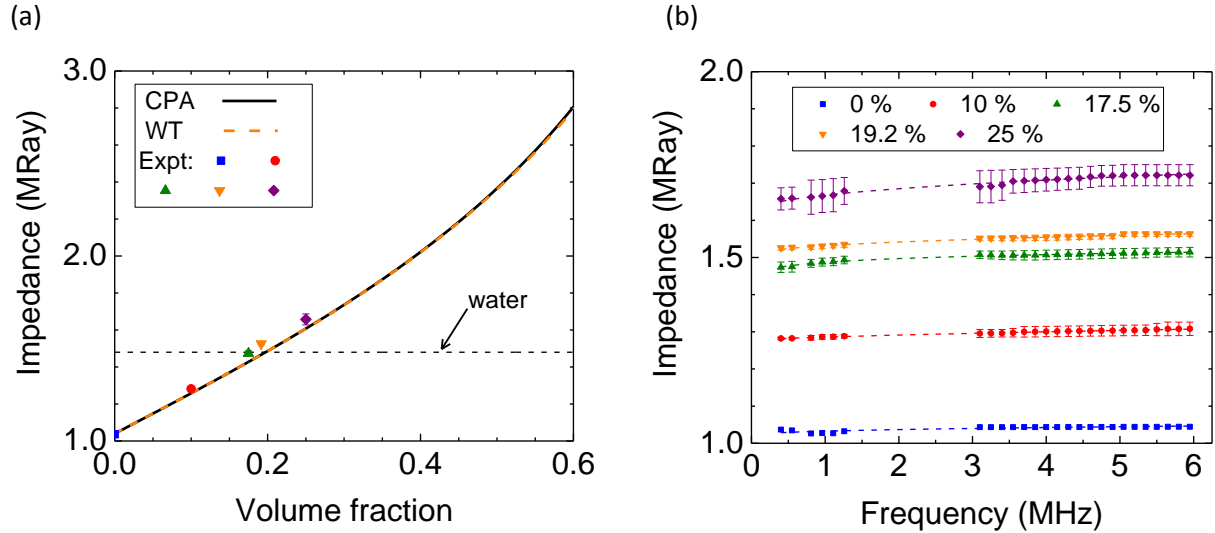

**Fig. 4.** (a) Impedance predicted by the two models and measured experimentally at 500 kHz. The dashed line represents the impedance of water at 20 °C. (b) Impedance over a wide range of frequencies for different concentrations of $TiO_2$ particles. Error bars represent the standard deviation of two different measurements on the same samples. Dashed lines are guides to the eye.

Figure 5a shows the time signals probed after a short pulse (centered at 500 kHz) was reflected from pure PDMS (blue line) and from PDMS loaded with 17.5 % of $TiO_2$ particles (red line). In the pure PDMS case, we note that two separate echoes are seen arising from the reflections at both water-PDMS interfaces. However, almost no reflection is observed in the second case (17.5 % of $TiO_2$ particles), demonstrating that excellent impedance matching is achieved over a wide frequency range. It is also instructive to visually inspect the transmitted signals, which are shown in Figure 5b; this figure clearly indicates an increasing time delay (associated with a slowing down of the sound velocity) between the signals that propagated through water, PDMS and PDMS-$TiO_2$, respectively. Including $TiO_2$ particles also has the effect of increasing the attenuation in PDMS, as can also be seen in Figure 5b, where the amplitude of the transmitted signal is smaller in the case of the PDMS-$TiO_2$ sample than in the case of pure PDMS.

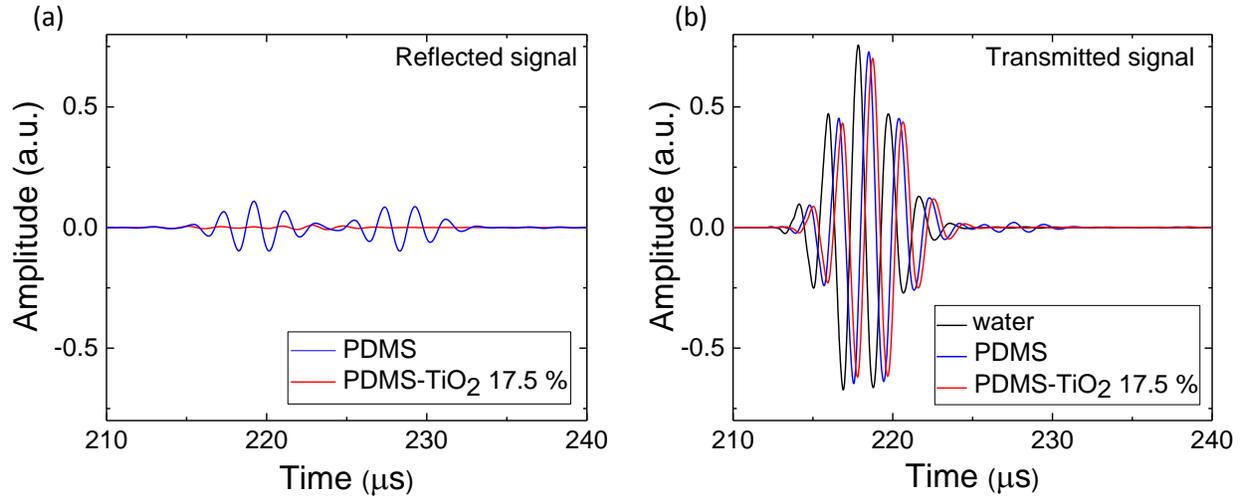

**Fig. 5.** Typical signals recorded for a sample of pure PDMS (blue) and PDMS with 17.5 % of TiO$_2$ particles (red). To facilitate comparison between the reflected (a) and transmitted (b) signals, the same vertical scale is used in both figures. The experiment was done with 500 kHz transducers (Gaussian pulse, with a 6 dB bandwidth of 200 kHz). In (b), the pulse through PDMS was shifted in time to compensate for the difference in thickness of this sample relative to the PDMS-TiO$_2$ sample, so that the arrival times of the pulses are consistent with each other and correspond to a sample thickness of 2.16 mm. The black line is the reference signal through water.

By analyzing such transmission measurements quantitatively, as described at the end of section 3, the evolution of the attenuation with frequency and with concentration of TiO$_2$ can be investigated (Figure 6). As shown in figure 6a, the frequency dependence is well described by a power law with exponent 1.45, $\alpha \propto f^{1.45 \pm 0.10}$, so that that at high frequencies, as usual, the signals are more attenuated. Figure 6b shows that the attenuation increases approximately linearly with the volume fraction of TiO$_2$ particles, and can become quite substantial at high frequencies and particle concentrations. Indeed, the attenuation is higher than the predictions of either model, suggesting that viscous losses not included in these theoretical calculations are important. These viscous losses are presumably associated with the motion of hard inclusions in the PDMS viscoelastic matrix, and would require in-depth modeling beyond the scope of the present work. In the context of this paper, the relevance of this enhanced attenuation is that it can facilitate the development of some of the applications of impedance-matched PDMS-TiO$_2$ that are mentioned in the next section.

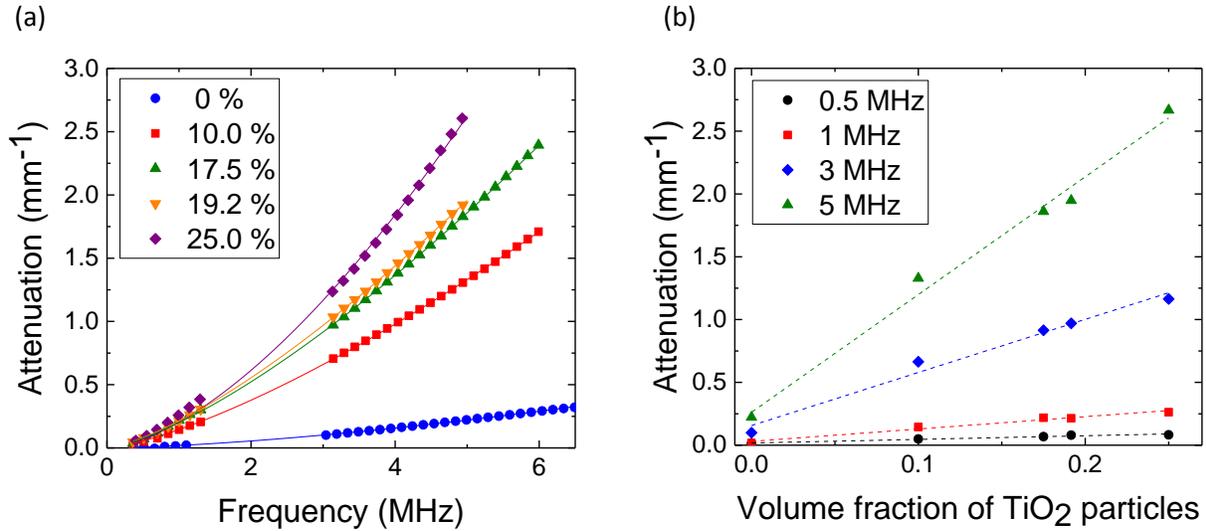

**Fig. 6.** (a) Attenuation as a function of frequency for the different particle concentrations and (b) attenuation at representative frequencies as a function of particle volume fraction. Data in (a) are fitted with the power law $\alpha \propto f^{1.45 \pm 0.10}$ (solid lines). The dashed lines in (b) indicate that the increase in attenuation with TiO$_2$ particle concentration is approximately linear.

## 5. Conclusions

We have demonstrated the possibility of tuning the impedance of PDMS by loading it with TiO$_2$ particles, and reported an experimental realization of almost perfect impedance matching with water. Since the impedance variation with frequency is extremely small, the impedance matching is robust over a wide range of frequencies, enabling the efficient use of this material for broadband pulsed experiments. The Waterman & Truell and Coherent Potential Approximation models can both be used to estimate the concentration of particles needed.

The measured attenuation depends on the frequency and concentration of particles, and adds a feature to this composite material that can be advantageous for some possible applications. For example, this material could be ideal as the matrix for creating ultrathin perfectly absorbing metalayers of bubbles [3], as the impedance matching properties will remove all the unwanted reflections at the interface between the layer and the surrounding water, and the attenuation will make the absorption of waves even more efficient. Another interesting application for this material is the realization of Perfectly Matched Layers (PML) that can be very useful to avoid standing wave effects. Indeed, one can imagine fully dissipating an incoming wave with a slab of this PDMS-$TiO_2$ material which would have a thickness adapted to its effectiveness in the targeted frequency range.

The absence of reflection of waves incident on the water/PDMS-$TiO_2$ interface makes this material very useful for designing sample cells. Because PDMS curing can be achieved in a mold, our composite material can be very easily shaped, which makes it very easy to handle and use. PDMS-$TiO_2$ can therefore be used for cells with nearly acoustically transparent walls to contain dispersions of particles, emulsions, or bubbly liquids, so that they can be more easily probed with ultrasound.

**Acknowledgements**

Support from the Natural Sciences and Engineering Research Council of Canada (NSERC) is gratefully acknowledged.

**References**

[1] T.G. Álvarez-Arenas, Acoustic impedance matching of piezoelectric transducers to the air, IEEE transactions on ultrasonics, ferroelectrics, and frequency control 51(5) (2004) 624-633.

[2] J.H. Goll, B.A. Auld, Multilayer impedance matching schemes for broadbanding of water loaded piezoelectric transducers and high Q electric resonators, IEEE Transactions on Sonics and Ultrasonics, 22(1) (1975) 52-53.

[3] V. Leroy, A. Strybulevych, M. Lanoy, F. Lemoult, A. Tourin, J.H. Page, Superabsorption of acoustic waves with bubble metascreens, Physical Review B 91(2) (2015) 020301.

[4] J.B. Pendry, Negative refraction makes a perfect lens, Physical Review Letters 85(18) (2000) 3966.

[5] P. Sheng, Introduction to wave scattering, localization and mesoscopic phenomena, 2nd edition, Springer-Verlag Berlin Heidelberg, 2006.

[6] E.H. Kerner, The elastic and thermo-elastic properties of composite media, Proceedings of the physical society Section B 69(8) (1956) 808.

[7] Z. Hashin, S. Shtrikman, Note on a variational approach to the theory of composite elastic materials, Journal of the Franklin Institute 271(4) (1961) 336-341.

[8] R.M. Christensen, K.H. Lo, Solutions for effective shear properties in three phase sphere and cylinder models, Journal of the Mechanics and Physics of Solids 27(4) (1979) 315-330.

[9] P.C. Waterman, R. Truell, Multiple scattering of waves, Journal of Mathematical Physics 2(4) (1961) 512-537.

[10] P.S. Epstein, R.R. Carhart, The Absorption of Sound in Suspensions and Emulsions. I. Water Fog in Air, J. Acoust. Soc. Am. 25 (1953) 553.

[11] J.R. Allegra, S.A. Hawley, Attenuation of Sound in Suspensions and Emulsions: Theory and Experiments, J. Acoust. Soc. Am. 51 (1972) 1545-1564.

[12] P.Y. Longin, C. Verdier, M. Piau, Dynamic shear rheology of high molecular weight polydimethylsiloxanes: comparison of rheometry and ultrasound, Journal of Non-Newtonian Fluid Mechanics 76(1) (1998) 213-232.

[13] V. Leroy, K.M. Pitura, M.G. Scanlon, J.H. Page (2010), The complex shear modulus of dough over a wide frequency range, Journal of Non-Newtonian Fluid Mechanics 165(9) (2010) 475-478.

[14] D.H. Chung, W.R. Buessem, The Voigt-Reuss-Hill (VRH) Approximation and the Elastic Moduli of Polycrystalline ZnO, TiO2 (Rutile), and α-Al2O3, Journal of Applied Physics 39(6) (1968) 2777-2782.

[15] A. Strybulevych, V. Leroy, M. G. Scanlon, J.H. Page, Characterizing a model food gel containing bubbles and solid inclusions using ultrasound, Soft Matter 3 (2007), 1388-1394.

[16] For example, if $\alpha v/2\omega = 0.02$, the complex impedance given by Equ. (1) differs in magnitude and phase from the approximate value, Equ. (10), by 0.02% and $1°$, respectively.

[17] R.B. Lindsay, R.T. Beyer, Acoustics, in: H. L Anderson (Editor in Chief), A physicists desk reference: The second edition of physics vade mecum. American Institute of Physics New York, 1989, pp. 52-64.